\documentclass[10pt]{article}
\usepackage[utf8x]{inputenc}
\usepackage[english]{babel} % английские переносы
\usepackage{graphicx}          % для включения графических изображений
\usepackage{cite}              % для корректного оформления литературы
\usepackage{xcolor}

\usepackage{amsmath}
\begin{document}

523.4-52: 573.552
\begin{center}
Analysis of Methods for Computing the Trajectories of Dust Particles
in a Gas–Dust Circumstellar Disk
\end{center}

\begin{center}
O.P.~Stoyanovskaya\footnote{Boreskov Institute of Catalysis, Lavrentieva, 5, 630090, Novosibirsk, Russia, stop@catalysis.ru}, E.I.~Vorobyov\footnote{Institute of Astrophysics, University of Vienna, Vienna, Austria, eduard.vorobiev@univie.ac.at}, V.N.~Snytnikov\footnote{Boreskov Institute of Catalysis, Lavrentieva, 5, 630090, Novosibirsk, Russia, snyt@catalysis.ru} 

\end{center}
% Аннотация заключается в окружение abstract
\sloppy
\begin{abstract}

A systematic analysis of methods for computing the trajectories of solid-phase particles applied in modern astrophysics codes designed for modeling gas–dust circumstellar disks has been carried out for the first time. The motion of grains whose velocities are determined mainly by the gas drag, that is, for which the stopping time or relaxation time for the velocity of the dust to the velocity of the gas $t_{stop}$ is less than or comparable to the rotation period, are considered. The methods are analyzed from the point of view of their suitability for computing the motions of small bodies, including dust grains less than 1 $\mu m$ in size, which are strongly coupled to the gas. Two test problems are with analytical solutions. Fast first order accurate methods that make it possible to avoid additional restrictions on the time step size $\tau$ due to gas drag in computations of the motion of grains of any size are presented. For the conditions of a circumstellar disk, the error in the velocity computations obtained when using some stable methods becomes unacceptably large when the time step size is $\tau > t_{stop}$. For the radial migration of bodies that exhibit drifts along nearly Keplerian orbits, an asymptotic approximation, sometimes called the short friction time approximation or drift flux model, gives a relative error for the radial-velocity computations equals to $St^2$ where St is the Stokes number, the ratio of the stopping time of the body to some fraction of the rotation period (dynamical time scale) in the disk. 

\end{abstract}
\vspace{\baselineskip}

\textbf{DOI}: 10.1134/S1063772917120071

\vspace{\baselineskip}
Keywords: protoplanetary discs, two-phase medium, two-fluid approach, splitting method
% \section{название} - заголовок раздела первого уровня
% \subsection{название} - заголовок раздела второго уровня
% \subsubsection{название} - заголовок раздела третьего уровня
% Не используйте уровень вложенности заголовков больше трех!
% Каждый абзац текста в статье начинается командой \par или пустой
% строкой.

\section{Introduction}
Molecular clouds, in which stars form with their circumstellar disks, are comprised 98-99\% of hydrogen and helium. All the remaining elements are contained in solid-phase dust particles, some of which go into forming planets in the circumstellar disks. The fraction of solid-phase particles in these disks relative to hydrogen and helium can change appreciably in time and space. In the process of forming planets from dust particles, bodies of larger sizes form along the way, right to multi-kilometer planetesimals and protoplanets. During the migration of bodies through a disk, the growth and fragmentation of solid-phase particles can change the gas-dynamical regimes for the interaction of individual bodies and dust particles with the ambient gas. The change in these regimes from free-molecular interactions with the gas flowing around bodies to the gas acting as an approximately continuous medium is especially important for modeling accreting protoplanetary disks on their early massive stages associated with the growth of dust particles.

Currently, modeling of processes in gas–dust disks, especially two-phase circumstellar disks, is usually based on a gas-dynamical approximation (see, for example \cite{HaworthEtAl2016,ZiglinaMakalkin2016}). In this approximation, with a two-fluid approach, the dynamics of dust grain cloud are calculated separately from the gas dynamics using gas-dynamical equations (e.g. \cite{ZhuDust,ChaNayakshinDust2011,RiceEtAl2004,BaiStone2010ApJS,FranceDustCode,Pignatale2016}). In a number of cases, it is convenient to represent the mean velocity of the medium and separately the relative velocity between the gas and bodies using a gas-dynamical system of equations for a two-phase medium of dust and gas (e.g., \cite{BateDust2014,LaibePrice2014OneFluidDust}). This is sometimes called a one-fluid approach. The initial systems of equations for one-fluid and two-fluid approaches are mathematically equivalent (e.g., \cite{LaibePrice2014OneFluidDust}), and the way of describing the system is determined by the chosen numerical method.

We note separately that, in a number of astrophysical problems, it is possible to consider particles with different velocities located at one point in space. In other words, an intersection of particle trajectories is considered in the mathematical model for the dust subsystem. Lagrangian–Eulerian (e.g., Particle-in-Cell \cite{Hockney}) and fully Lagrangian (e.g., Smoothed Particle Hydrodynamics \cite{MonaghanKocharyan1995} or the method of Osiptsov \textcolor{red}{cite{13}}) have been developed for such problems. Such approaches can be used to simulate the collective dynamics of large bodies in circumstellar disks (e.g. \cite{1MNRAS}).

All these approaches for a two-phase medium re-
quire the integration over time of the equation for the
velocity of the bodies or the relative velocity of the
gas and bodies, which is influenced by aerodynamical
drag, gravitation, and other forces.

A computational difficulty arises when the typical time for the aerodynamical forces (the time over which the velocity of the dust relative to the gas is appreciably changed by drag between the dust and gas) is several orders of magnitude shorter than the typical time for the action of other forces (other names for the former time scale are the settling time, time for relaxation of the dust velocity to the gas velocity, and the stopping time). For explicit methods, the presence of several time scales in the system means that the time step must be smaller than the minimum typical time. However, the integration must be carried out over an interval exceeding several times the maximum typical time scale. For problems where each time step is computationally expensive (two- and three dimensional equations, taking into account self-gravitational or magnetic fields together with other physical–chemical processes), explicit methods for the integration of the dynamics of small dust consume unacceptably high computational resources. 

One approach to resolving this problem is a transition to an asymptotic approximation \cite{JohansenKlahr2005}, known as the short friction time approximation or drift flux model. In this approximation, the velocities of the bodies and gas are linked by a simple algebraic relation, which we will derive in Section \ref{sec:ShortFriction}. Section \ref{sec:approxBoundary} presents a derivation of the necessary conditions for applicability of this approximation, which demonstrates that it provides correct results only for dust of a limited size.

On the other hand, the simulation results (e.g. \cite{BrauerDullemondHenning2008}) and observations indicate the growth of dust sizes from 1 $\mu m$ to 1-100 cm or more over the first 10 million years of the evolution of a circumstellar disk. 

Therefore, universal algorithms enabling the integration of the equations of motion for dust with sizes from micrometers to 1–10 m are of interest when modeling the dynamics of a gas–dust disk over long time scales. Note that calculation of the orbits of larger bodies, whose velocities are determined mainly by gravitation, rather than gas drag, requires high-order accurate methods, able to reproduce the trajectories of the bodies over a large number of orbits. Here, we limit our consideration to numerical methods for bodies for which the times for the relaxation of the dust velocity relative to the gas velocity $t_{stop}$ are shorter than or comparable to the orbital period (dynamical time scale). We first present a systematic analysis of the numerical schemes for computing the trajectories of dust particles applied in modern astrophysical problems. We have obtained for the first time expressions for the local errors in the approximations used in the schemes considered as a function of the parameters of the problem and the computational path, and determined theoretically the orders of these approximations. We present recommendations for choosing methods for computing the trajectories of dust particles based on our results.

The paper has the following structure. Section ~\ref{sec:drag_regimes} presents the regimes for aerodynamical friction in a circumstellar disk. Section ~\ref{Schemes} describes the general approaches used to construct fast numerical schemes. The methods we tested are presented in Section ~\ref{TestSection}.  Sections ~\ref{sec:dustybox} and \ref{sec:quasianalytics} describe test problems with analytical solutions, and the results of estimating the actual accuracy of these methods are presented in
Sections ~\ref{sec:DustyboxResults} and \ref{sec:nonlinear_results}. Section \ref{sec:approxBoundary} presents a derivation of the necessary condition for applicability of the short friction time approximation, and Section ~\ref{sec:resume} summarizes our results.

\section{Regimes of aerodynamical drag in a circumstellar disk}
\label{sec:drag_regimes}
The aerodynamical force exerted by a gas with density $\rho$ on a spherical body with radius $a$ has the form

\begin{equation}
\label{Drag}
    \begin{bf} F_{\rm drag} \end{bf} = -\frac{1}{2} C_D \pi a^2 \rho \|\bf{u}-\bf{v}\| (\bf{u}-\bf{v}),
\end{equation}
where $u$ is the velocity of the gas and $v$ is the velocity of the body. $C_D$ is the dimensionless drag coefficient, whose value depends on the size of the body and the parameters for flow around the body \cite{Weidenschilling1977}: 

\begin{equation}
\label{eq:Drag_Coefficient}
C_{D} = \begin{cases}
        \displaystyle \frac{8c_s}{3\|\bf{u}-\bf{v}\|} & a < \displaystyle \frac{9}{4} \lambda, \textrm {Epstein regime},\\
        24 R_{e}^{-1} & R_{e} < 1, \textrm {linear Stokes regime},\\
        24 R_{e}^{-0.6} & 1 < R_e < 800,\\
        0.44 & R_{e} > 800,
        \end{cases}
\end{equation}
where $\lambda$ is the mean free
path of molecules in the gas and $R_e$ is the Reynolds number, $c_s$ is the sound speed in the gas.

\begin{figure*}
    \includegraphics[scale=0.35]{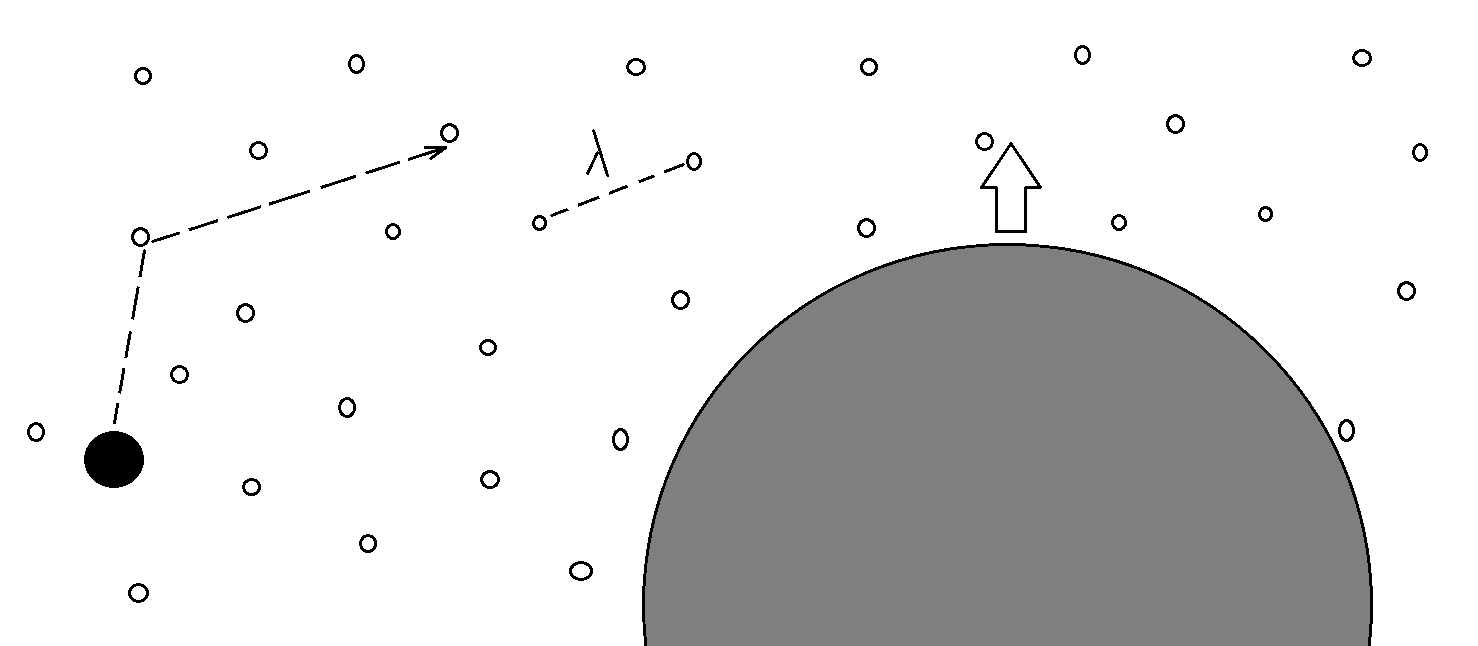} %[width=150mm]{fig2.eps}
  \caption{Schemes for gas flowing around the body. The left panel shows a body with radius $a < \displaystyle\frac{9}{4} \lambda$ undergoing collisions with individual gas molecules in the Epstein regime, and the left panel gas behaving like a continuous medium flowing around a body with radius $a > \displaystyle\frac{9}{4} \lambda$ in the Stokes regime.}
\label{fig:Regimes_Scheme}
\end{figure*}

Bodies whose radii are 
\begin{equation}
\label{eq:a_limit}
a < \displaystyle\frac{9}{4} \lambda,
\end{equation}
undergo collisions with individual molecules of the gas. This is called the Epstein or free molecular flow regime. Larger bodies interact with the gas as if it were a continuous medium in the Stokes regime. For such bodies, the drag coefficient depends on the Reynolds number $R_e$. A schematic of interactions of bodies with gas in the Epstein and Stokes regimes is presented in Fig. \ref{fig:Regimes_Scheme}. 

\begin{figure*}
    \includegraphics[scale=0.55]{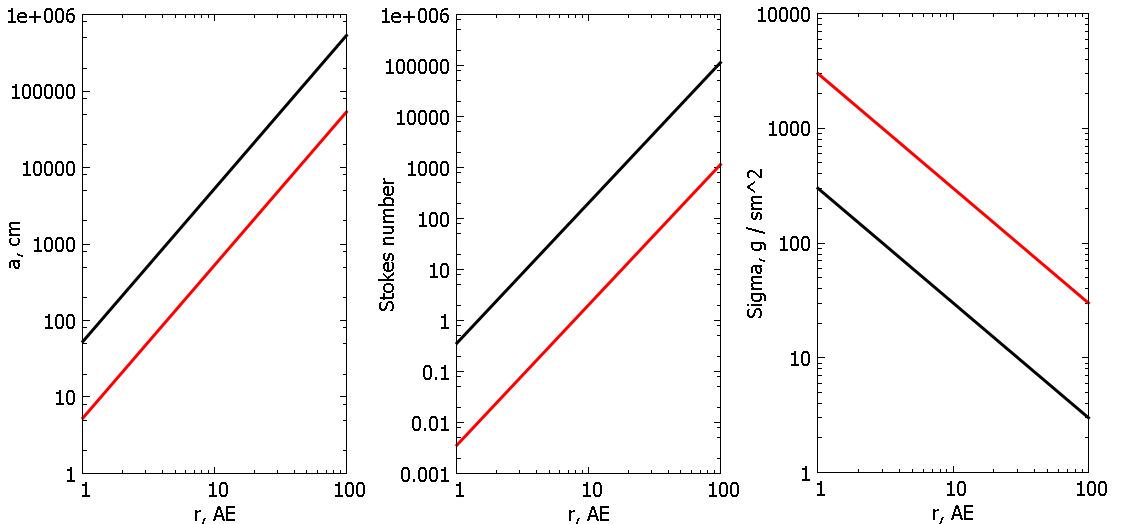} %[width=150mm]{fig2.eps}
  \caption{The left panel shows the maximum radius of a solid particle that can interact with gas in the Epstein regime, the central panel the maximum Stokes number characterizing the interaction of gas and bodies in the Epstein regime, and the right panel the surface density of a disk. The black line corresponds to a disk with mass $0.02 M_{\odot}$ and the red line to a disk with mass $0.2 M_{\odot}$.}
\label{fig:MaxStokes}
\end{figure*}

Let us now estimate the size of a body that can be described using the Epstein regime in a circumstellar disk. Following \cite{Vorobyov2010}, we adopted the characteristic dependence for the disk surface density on radius in the disc $\Sigma=\Sigma_0\left( \displaystyle \frac{r}{r_0} \right)^{-1}$ ($\Sigma = \int^{H}_{-H} \rho dz$, where $\rho$ is the gas volume density in the disk and H the disk height). Let us consider a disk that extends from 1 to 100 AU. We then obtain $\Sigma_0=300$~g/cm$^{2}$ for $r_0=10$ AU for a disk mass of $M_{\rm disc}=0.2M_{\odot}$ and $\Sigma_0=30$~g/cm$^{2}$ for $r_0=10$ AU for a disk mass of $M_{\rm disc}=0.02M_{\odot}$. The resulting surface-density profiles are presented in the right panel of Fig. \ref{fig:MaxStokes}. 

The mean-free path of a gas molecule is determined by $\lambda=\displaystyle \frac{m_{\rm H_2}}{\rho \sigma}$, where $m_{\rm H_2}$ is the mass of a hydrogen molecule and $\sigma$ is the cross section for elastic collisions between hydrogen molecules. Setting $H=0.1r$, $\rho=\displaystyle\frac{\Sigma}{H}$, and using the values $m_{\rm H_2}=3.32 \times 10^{-24}$~g and $\sigma=7 \times 10^{-16}$~cm$^2$ we can use (\ref{eq:a_limit}) to determine the maximum size of particles in the disk that interact with the gas in the free-molecular flow (Epstein) regime. The derived particle sizes at various disk radii are presented in the left panel of Fig.\ref{fig:MaxStokes}. We can see that bodies with radii less than 1~m  interact with the ambient gas in the Epstein regime in the outer part of an axially symmetric disk (distances of more than 10 AU from the protostar). Moreover, the formation of self-gravitating clumps at a radius of 100 AU whose densities exceed the background density by more than a factor of 100 makes it possible to use the Epstein regime as aerodynamical drag for bodies with sizes up to $\approx$1~m. Thus, in the outer part of a disk, the flow of gas around solid bodies with sizes from those of grains to those of boulders (and in the case of low-mass disks, to the sizes of planetesimals) can be described in a free-molecular flow (Epstein) regime. In this approach, the acceleration $\bf g_{\rm drag}$ experienced by the body interacting with the gas can be written
\begin{equation}
\label{eq:g_drag_Epstain}
\begin{bf} g_{\rm drag} \end{bf}=\frac{\bf{u}-\bf{v}}{t_{\rm stop}},
\end{equation}
where $t_{\rm stop}$ is the stopping time of the particle, which is given by
\begin{equation}
\label{eq:t_stop_definition}
t_{\rm stop}=\displaystyle\frac{m_{\rm s}\|\bf{u}-\bf{v}\|}{\|\begin{bf}F_{\rm drag} \end{bf}\|},
\end{equation}
and $m_{\rm s}$ is the mass of the particle, $m_{\rm s}=4 \pi \rho_{\rm s} a^3/3$ where $\rho_s$ is the density of the substance of which the dust consists. We obtain by virtue of (\ref{Drag}) and (\ref{eq:Drag_Coefficient}) 
\begin{equation}
\label{eq:t_stop_rho}
t_{\rm stop}=\frac{a \rho_{\rm s}}{\rho c_s}.
\end{equation}
For a circumstellar disk $\displaystyle\frac{c_{\rm s}}{v_{\rm K}}=\frac{H}{r}$ where $v_{\rm K}=\displaystyle \sqrt{\frac{GM}{r}}$ is the Keplerian velocity, $M$ the mass of the star, and $G$ the gravitational constant. Then,
\begin{equation}
\label{eq:t_stop_sigma}
t_{\rm stop}=\frac{a \rho_{\rm s}}{\Sigma \Omega_{K}}, 
\end{equation}
where $\displaystyle\Omega_K = \frac{v_K}{r}$ is the Keplerian frequency, which characterizes the dynamical time scale for the motion
of a particle in a Keplerian orbit in the circumstellar disk.

We characterized the coupling of dust to gas in a circumstellar disk due to aerodynamical drag using the Stokes number or dimensionless stopping time:
\begin{equation}
\label{eq:Stokes}
St=t_{\rm stop} \Omega_K,
\end{equation}
which takes the form $St= \displaystyle\frac{a \rho_{\rm s}}{\Sigma}$ in the Epstein regime. Thus, the smaller the radius of a body,the stronger it is tied to the gas ($St<1$). Setting $\rho_{\rm s}=2.2$~g/cm$^{3}$ and using (\ref{eq:a_limit}) we can estimate the maximum Stokes number for a body in a circumstellar disk interacting with the gas in the Epstein regime. The resulting values are presented in the central panel of Fig.\ref{fig:MaxStokes}; we can see that, under the disk conditions considered, bodies interacting with the gas in a free-molecular flow regime can both experience strong adhesion with the gas ($St \ll 1$), and move essentially independently of the gas ($St \gg 1$). Note that $St \ll 1$ does not mean that the dust velocity coincides with the gas velocity, so that there is an absence of dust drift.

\section{Methods for computing the dust velocity}
\label{Schemes}

To understand the requirements for numerical algorithms used to compute the dynamics of solid bodies in a disk, we considered a model equation of motion for a single body in a medium with friction. For simplicity, we analyzed the equation for one velocity component (the transition to a system of ordinary differential equations for the three velocity components is straightforward):

\begin{equation}
\label{eq:gen_motion}
\frac{dv}{dt}=g+g_{\rm drag}, 
\end{equation}
where $g_{\rm drag}=\displaystyle\frac{u-v}{t_{\rm stop}}$ is the acceleration of the body due to friction in the gas, $g$ is its acceleration due to non-aerodynamical forces, and $u(t)$ is the gas velocity as a function of time. An explicit first order accurate method for this equation has the form
\begin{equation}
\label{eq:Euler}
        \displaystyle \frac{v^{n+1}-v^{n}}{\tau} = g^n+\displaystyle\frac{u^n-v^n}{t_{\rm stop}},      
\end{equation}
where $v^n$ is the velocity of the body at the current time $t$ (in time step $n$) and $v^{n+1}$ the velocity of the body at the following time $t+\tau$ (in time step $n+1$). If $g^n=0$ and $u^n=0$, Eq. (\ref{eq:Euler}) acquires the form
\begin{equation}
\label{eq:Euler2}
        v^{n+1}=\left(1-\displaystyle\frac{\tau}{t_{\rm stop}} \right)v^n=\left(1-\displaystyle\frac{\tau}{t_{\rm stop}} \right)^n v_0.      
\end{equation}
It follows from (\ref{eq:Euler2}) that for $v^{n+1}$ to be finite for any $n$ a necessary condition is that: $\left|1-\displaystyle\frac{\tau}{t_{\rm stop}} \right|<1$, which is equivalent to the form
\begin{equation}
\label{eq:tauEuler2}
  \tau<2t_{\rm stop}.      
\end{equation}
%{\bf Почему не $\tau<t_{\rm stop}$, при котором скорость обращается в ноль?}

Let us consider the effect of this constraint on the time step size. Suppose a grain has a radius of $1$~$\mu m$; for a massive disk in which $\Sigma=100$~g/cm$^{2}$ at $r=20$ AU, we find from (\ref{eq:t_stop_sigma}) $t_{\rm stop} \approx 100$~s. 

When integrating the equations of the disk dynamics (usually over ten or more orbital periods of the outer part of the disk), the time step size is determined from the Courant condition $\tau<\displaystyle\frac{\Delta r}{v}$, where $\Delta r$ is the step for discretization of the spatial derivatives (the size of a grid cell or smoothing length in Smoothed Particle Hydrodynamics). Let us estimate the typical size of this step. Letting the inner disk radius be $r=1$~AU, we carried out simulations of the disk dynamics in cylindrical coordinates, dividing a total rotation into 256 cells. This yields
\begin{equation}
\label{eq:tauKurant}
\tau=\displaystyle\frac{\Delta r}{v_{\rm K}}=\frac{2\pi r}{256 v_{\rm K}} \approx 1,23 \times 10^5 \ \textrm{s}. 
\end{equation}
Satisfying the condition (\ref{eq:tauEuler2}) requires computations using a factor of 1000 more time steps. Experience with solving stiff problems indicates that constraints on the time step size can be eased appreciably if an implicit scheme is used for the computations, such as the simplest version of the implicit first order accurate scheme:
\begin{equation}
\displaystyle \frac{v^{n+1}-v^{n}}{\tau} = g^{n+1}+\displaystyle\frac{u^{n+1}-v^{n+1}}{t_{\rm stop}}.   
\end{equation}
In the general case, $g^{n+1}, u^{n+1}$, and $v^{n+1}$ are unknown, and computing them using such a scheme requires iterations. In the case of the motion of a body in a plane or in space, a matrix inversion is required at each iteration. Therefore, modern simulations of the dynamics of gas–dust disks have tended to use the faster approaches considered below. Section \ref{sec:ShortFriction} is concerned with an approach involving simplification of the initial problem and Section \ref{sec:Zhu} with numerical methods for solving the entire problem.

\subsection{Short friction time approximation}
\label{sec:ShortFriction}
Let us suppose that we require the solution of a system of equations of motion of a grain with velocity $v$ in gas with velocity $u$. We considered the case when the gas influences the dust, but, due to its low mass concentration, the dust does not affect the velocity of the gas:  
\begin{equation}
\label{eq:simptote}
\left\{
 \begin{array}{lcl}
        \displaystyle
        \frac{d v}{dt} = g + \frac{u-v}{t_{\rm stop}},\\
        \displaystyle
        \frac{d u}{dt}= g_u. \\
    \end{array}
\right.
\end{equation}
where $g_u$ is the acceleration acting on a gaseous volume in the vicinity of a solid body. Subtracting the  second equation from the first and multiplying by the Stokes number $St$ yields the equivalent system
\begin{equation}
\label{eq:simptoteSmallParameter}
\left\{
 \begin{array}{lcl}
        \displaystyle
        St\frac{d (v-u)}{dt} = St (g-g_u) + \Omega (u-v),\\
        \displaystyle
        \frac{d u}{dt}= g_u. \\
    \end{array}
\right.
\end{equation}
When $St \ll 1$ (a small body and a large difference between the dynamical time and the stopping time), the first equation of this system has the solution $u \approx v$ with accuracy to within $St$. The subsequent approximation is obtained if we take into account that $St \displaystyle \frac{d(u -v)}{dt}$ is of second order of smallness in $St$ relative to the other terms. Therefore, we have to within first-order accuracy in $St$
\begin{equation}
\label{eq:simptoteDegenerate}
\left\{
 \begin{array}{lcl}
        St (g-g_u) + \Omega (u-v)=0,\\
        \displaystyle
        \frac{d u}{dt}= g_u. \\
    \end{array}
\right.
\end{equation}
It follows that
\begin{equation}
\label{eq:SFTA}
v=g _{\rm rel} t_{\rm stop}+u,
\end{equation}
where $g_{\rm rel}=g-g_u$ is the difference in the accelerations acting on the gas and on the body.

The physical meaning of Eq. (\ref{eq:SFTA}) is the use of the <<steady-state>> velocity of the particles relative to the gas in the computations, that is, the use of a quantity constructed based on the assumption that the relative velocity between the gas and bodies is constant when the forces are constant. This method for computing the velocity of a small grain is economical and natural for complex disk models, and was applied in \cite{JohansenKlahr2005,ZhuDust}. The necessary condition for applicability of this approximation for particles migrating along nearly Keplerian orbits is derived in Section \ref{sec:approxBoundary}. 

\subsection{Methods for Integrating the Entire Problem}
\label{sec:Zhu}
\subsubsection{Regularization}
The idea behind this method is replacing the explicit computation of the aerodynamical acceleration by the regularized acceleration  $g^n_{\rm drag}=\displaystyle\frac{u^n-v^n}{t_{\rm stop}+\tau}$. This regularization method is used, for example, in \cite{ZhuDust}. We will show in Section \ref{sec:approximation} that a correct choice of the numerical scheme can make it possible to obtain a correct asymptotic  radial drift velocity for bodies of arbitrarily $t_{\rm{stop}}$. 

\subsubsection{Quasianalytical integration}
\label{Snyt}
The idea behind this method is that the stiff term in the equation of motion depends linearly on the velocity of the dust relative to the gas. The velocity of a grain at a time $\tau$ can then be calculated using a finite-difference method, and we have from the analytical solution of the Cauchy problem for a linear equation
\begin{equation}
\label{DragStageAnalyt}
\left\{
 \begin{array}{lcl}
        \displaystyle 
        \frac{d v}{dt} = g^{num} + \frac{u^{num}-v}{t_{\rm stop}},\\
        v|_{t=0} = v^n. \\
    \end{array}
\right.
\end{equation}
where $g^{num},u^{num}$ are constants. Substituting in the system (\ref{DragStageAnalyt}) $g^{num}=g^n$ and $u^{num}=u^n$ then yields
\begin{equation}
\label{eq:quasianalytSolution}
v^{n+1}=(g^n t_{\rm{stop}}+u^n)+(v^n-g^n t_{\rm{stop}} -u^n)e^{-\displaystyle\frac{\tau}{t_{\rm{stop}}}}
\end{equation}

This approach was used, for example, in \cite{BateDust2014,BateDust2015, PanPadoan2013, Rosotti2016}. 

\subsubsection{Mixed layer scheme}
\label{Nayakshin}
In the mixed layer scheme, the aerodynamical acceleration is computed with the velocity of the body calculated using an implicit scheme in the new time step, and with the gas velocity taken from the previous time step: 
\begin{equation}
g^n_{\rm drag}=\frac{u^n-v^{n+1}}{t_{\rm stop}}.
\end{equation}
%Эта схема используется, например, в работе \cite{ChaNayakshinDust2011}.

Based on the approaches described, a large number of schemes can be constructed for Eq. (\ref{eq:gen_motion}) which will be unconditionally stable and provide constrained solutions for any $\tau$. Apart from one-step schemes, when the right-hand side of the equation is treated like a single operator, the operator splitting technique physical processes is often used in simulations of complex physical systems. We restricted our consideration to schemes that (a) have already been used in other computations and (b) can be realized with a minimum number of arithmetic operations. The aim of our study is to show that unconditionally stable schemes that have been applied, and which require the same number of arithmetic operations, can differ
substantially in the accuracy of obtained solutions. Another aim is to identify optimal methods from this point of view.

\section{Tested schemes}
\label{TestSection}
\subsection{Shord Friction Time Approximation}

An asymptotic value for the velocity of the particles relative to the gas is specified in the computations. In this approximation, the dust velocity is taken to be
\begin{equation}
\label{eq:SFTA}
v=g _{\rm rel} t_{\rm stop}+u,
\end{equation}
where $g_{\rm rel}=g-g_u$ is the difference of the accelerations acting on the gas and on the body.

\subsection{Mixed Layer Scheme}
\label{eq:Nayakshin}
This scheme has the form: 
\begin{equation}
\label{eq:Nayakshin}
\displaystyle\frac{v^{n+1}-v^n}{\tau}=\frac{u^n-v^{n+1}}{t_{\rm stop}}+g^n.
\end{equation}
This scheme was used, for example, in \cite{ChaNayakshinDust2011}.

\subsection{Quasianalytical Scheme without Operator Splitting}

Setting $g^{num}=g^n$ and $u^{num}=u^n$ in the system (\ref{DragStageAnalyt}) yields
\begin{equation}
\label{eq:quasianalytSolution}
v^{n+1}=(g^n t_{\rm{stop}}+u^n)+(v^n-g^n t_{\rm{stop}} -u^n)e^{-\displaystyle\frac{\tau}{t_{\rm{stop}}}}
\end{equation}
This scheme was used, for example, in \cite{NayakshinDust2017}.

\subsection{Schemes Based on Operator Splitting with Respect to Physical Processes}
We tested schemes based on this method, which are of first-order accuracy in the time(e.g., \cite{BateDust2015,ZhuDust}); the first stage of a separation scheme is finding the velocity of an individual body due to the nonaerodynamical (e.g., gravitational) acceleration,
\begin{equation}
\label{eq:GravStage}
        \displaystyle \frac{v^{n+1/2}-v^{n}}{\tau} = g,      
\end{equation}
and the second stage is correction of the grain velocity due the gas drag:
\begin{equation}
\label{eq:DragStage}
       \displaystyle\frac{v^{n+1}-v^{n+1/2}}{\tau} = g_{\rm drag}.           
\end{equation}

Stages (\ref{eq:GravStage}) and (\ref{eq:DragStage}) are not commutative; when their order is changed, we obtain the scheme

\begin{equation}
\label{eq:IndirectDragStage}
        \displaystyle \frac{v^{n+1/2}-v^{n}}{\tau} = g_{\rm drag},      
\end{equation}

\begin{equation}
\label{eq:IndirectGravStage}
       \displaystyle\frac{v^{n+1}-v^{n+1/2}}{\tau} = g.           
\end{equation}
Here, $v^{n+1/2}$ is the intermediate velocity of the grain. 
We will refer to (\ref{eq:GravStage}), (\ref{eq:DragStage}) s the scheme with direct operator order and to (\ref{eq:IndirectDragStage}), (\ref{eq:IndirectGravStage}) as the scheme with reverse operator order. 

\subsubsection{Regularization}
The regularization method with the direct operator order has the form 
\begin{equation}
\label{eq:directZhu}
\left\{
 \begin{array}{lcl}
        \displaystyle 
        \frac{v^{n+1/2}-v^n}{\tau} = g^n, \\
        \displaystyle
        \frac{v^{n+1}-v^{n+1/2}}{\tau} = \frac{u^n-v^{n+1/2}}{t_{\rm stop}+\tau},
  
    \end{array}
\right.
\end{equation}

and with the reverse operator order has the form
\begin{equation}
\label{eq:indirectZhu}
\left\{
 \begin{array}{lcl}
        \displaystyle 
        \frac{v^{n+1/2}-v^n}{\tau} = \frac{u^n-v^n}{t_{\rm stop}+\tau}, \\
        \displaystyle
        \frac{v^{n+1}-v^{n+1/2}}{\tau} = g^n.
  
    \end{array}
\right.
\end{equation}

A regularization method combined with an operator splitting technique was used, for example, in \cite{ZhuDust}. 

\subsubsection{Quasianalytical integration}

This method with the direct operator order has the form
\begin{equation}
\label{eq:directSnyt}
\left\{
 \begin{array}{lcl}
        \displaystyle 
        \frac{v^{n+1/2}-v^n}{\tau} = g^n, \\
        v^{n+1}=u^n+(v^{n+1/2}-u^n)\rm{e}^{-\displaystyle\frac{\tau}{t_{\rm stop}}},
          
    \end{array}
\right.
\end{equation}
and with the reverse operator order has the form
\begin{equation}
\label{eq:indirectSnyt}
\left\{
 \begin{array}{lcl}
      v^{n+1/2}=u^n+(v^n-u^n)\rm{e}^{-\displaystyle\frac{\tau}{t_{\rm stop}}},\\
      \displaystyle 
      \frac{v^{n+1}-v^{n+1/2}}{\tau} = g^n, \\  
    \end{array}
\right.
\end{equation}
Quasianalytical integration combined with operator splitting technique with respect to physical processes was used, for example, in \cite{BateDust2014,BateDust2015}. 

\section{Test 1. The DUSTYBOX equation and exact analytical solution}
\label{sec:dustybox}
We considered the slow migration of grains toward a protostar in a protoplanetary gaseous disk, whose pressure decreases with radius. We supposed that the gaseous disk was close to equilibrium, when its radial velocity is close to zero and its azimuthal velocity $u_\varphi$ is close to the Keplerian velocity $v_{\rm K}=\sqrt{\displaystyle\frac{GM}{r}}$, but does not reach this velocity. 

Let a dust particle move along its orbit with velocity $v_{\varphi}$. We assumed that the gravitational field acting on the particle is created purely by the protostar, not the disk. In this case, the particle experiences the radial acceleration $g=\displaystyle\frac{v_{\varphi}^2}{r} - \frac{MG}{r^2}$. If the motion is such that the radial velocity of a particle is much lower than the Keplerian velocity, the acceleration, which depends on the radius in the general case, can be taken to be constant. This leads to a model for the linear motion of a body under the action of this constant acceleration and the gas drag, which moves with the constant velocity $u$. This is a simplified model and does not decribe the transformation of the azimuthal velocity into radial velocity, but represents a non-trivial test for the numerical schemes (\ref{eq:Nayakshin}),(\ref{eq:directZhu}-\ref{eq:indirectSnyt}). This test is currently known as DUSTYBOX\cite{LaibePrice2011} and has been applied since 1995 \cite{MonaghanKocharyan1995}: 

\begin{equation}
\label{eq:straightlinemotion}
\left\{
 \begin{array}{lcl}
    
        \displaystyle 
        \frac{dv}{dt} = g+\frac{u-v}{t_{\rm stop}}, \\
        v|_{t=0} = v_0.
        
    \end{array}
\right.
\end{equation}
In our case, $v$ is the radial velocity of the migrating grains, $g$ the constant radial acceleration of the grains, and $u$ the radial velocity of the gas. The analytical solution of this problem has the form
\begin{equation}
\label{eq:simple_analytics}
v=(gt_{\rm stop}+u)+(v_0-g t_{\rm stop}-u){\rm e}^{-\displaystyle\frac{t}{t_{\rm stop}}}.
\end{equation}

It is clear that $v \rightarrow (g t_{\rm stop}+u)$ when $t \gg t_{\rm stop}$ , which coincides with the terminal velocity expressed by the
condition (\ref{eq:SFTA}).

\section{Results of testing the methods using DUSTYBOX}
\label{sec:DustyboxResults}

We took 100 dust particles with sizes from 1~$\mu m$ to 1~m and placed them at a distance $r=20$~AU from the protostar with the mass $M=1 M_\odot$. For simplicity, we assumed that $u=0$, and that $\Sigma(20 \rm AU)=100$~g/cm$^{2}$, with $g=-0.001\displaystyle\frac{v^2_{\rm K}}{r}$, $v_0=0.01v_{\rm K}$. 

For each of the particles, we carried out the integration (\ref{eq:straightlinemotion}) over 1000 rotations, i.e., over the time
interval $[0;T]=[0;2000 \pi \Omega_{\rm K}^{-1}]$, using a time step $\tau$ given by (\ref{eq:tauKurant}). Since $T \gg t_{\rm stop}$ for the entire range of particle sizes, this test can be used to estimate how close the computed stationary velocity is to the analytical value (\ref{eq:simple_analytics}). 

For convenience, all the constants needed to determine the parameters of the problem (\ref{eq:straightlinemotion}) are presented in Table \ref{tab:constants} and the parameters and initial conditions are presented in Table \ref{tab:params} in cgs units and the <<astronomical>> units indicated in the table.

\begin{figure*}
    \includegraphics[scale=0.7]{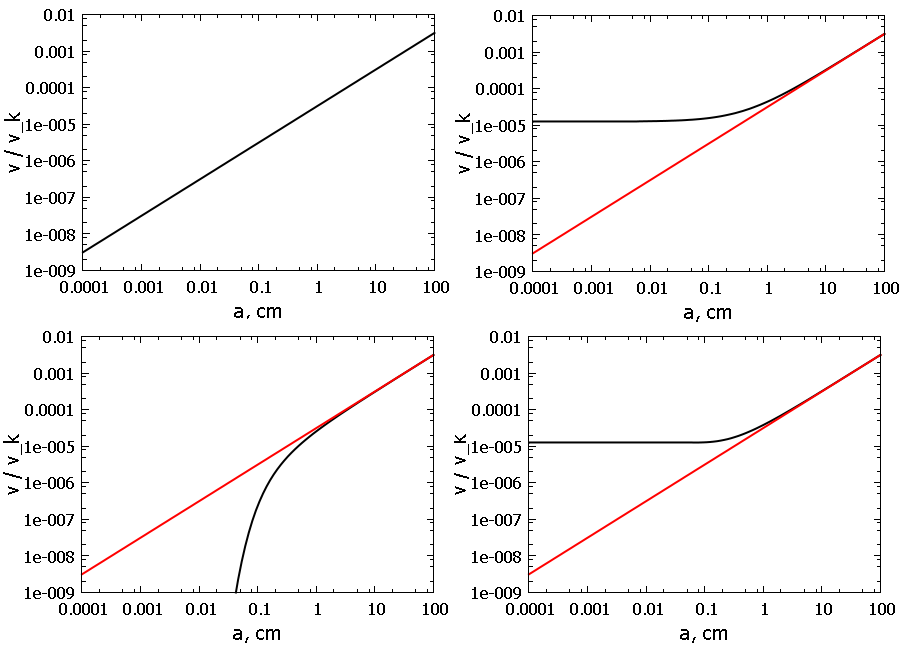} %[width=150mm]{fig2.eps}
  \caption{Numerical solutions of the problem (\ref{eq:straightlinemotion}). The upper row shows results for the regularization method and the lower row results for quasianalytical integration; the left panels show results for direct operator order (\ref{eq:directZhu}), (\ref{eq:directSnyt}) and the right panels
results for reverse operator order (\ref{eq:indirectZhu}), (\ref{eq:indirectSnyt}). The black curves show the numerical solutions and the red line the analytical solution.}
\label{fig:DustyBox_splitting}
\end{figure*}

The ratios of the computed speed $v$ and the Keplerian speed $v_{\rm K}$ at the radius $20$~AU are presented in Fig.\ref{fig:DustyBox_splitting}. Regularization with direct operator order (\ref{eq:directZhu}) produces radial migration rates close to the analytical values for all considered values of $t_{\rm stop}$. However, this same method with reverse operator order yields radial
migration rates for particles with sizes less than $1$~cm that substantially exceed the analytical values when $t_{\rm stop}<\tau$. It follows from Fig.\ref{fig:DustyBox_splitting} (upper right panel)
that, when using the scheme (\ref{eq:indirectZhu}) all bodies with
sizes less than $1$~cm experience the same gas drag and move in the disk like centimeter particles.

On the contrary, quasianalytical integration with direct operator order (\ref{eq:directSnyt}) appreciably underestimates the migration rates for bodies with sizes less than $1$~cm, essentially coupling their motion stiffly to the gas. Quasianalytical integration with reverse operator order (\ref{eq:indirectSnyt}) yields the same computational artefact as regularization with reverse operator order (\ref{eq:indirectZhu}), namely, an artificial enhancement in the migration rates of small bodies. We also found that the mixed layer scheme without operator splitting (\ref{eq:Nayakshin}) accurately reproduces the migration rates of any bodies; that is, the results essentially coincide with those of the scheme (\ref{eq:directZhu}) shown in the upper left panel of Fig.\ref{fig:DustyBox_splitting}. This test is not meaningful for the scheme (\ref{eq:quasianalytSolution}), since the numerical solution exactly coincides with the analytical solution when $g=const$.

\begin{table*}
\label{tab:constants}
\begin{minipage}{140mm}
\caption{Constants for determining the parameters of the
test problem (\ref{eq:straightlinemotion})}
\begin{center}
\label{tab:Nayakshin1}
\begin{tabular}{|c|c|}
\hline
   Quantity & Value in cgs units \\
    \hline
    $\rho_0$ & 2.992 $\times$ $10^{14}$  \\
  
   $a_{min}$ & $10^{-4}$  \\
 
   $a_{max}$ & $10^2$ \\
   
   $\rho_s$ & 2.2 \\
   
   $\Sigma(r_0)$ & 100 \\ 
   
   $M$ & 2 $\times$ $10^{33}$ \\
    
  \hline
\end{tabular}
\end{center}
\end{minipage}
\end{table*}

\begin{table*}
\label{tab:params}
\begin{minipage}{140mm}
\caption{Constants for determining the parameters of the
test problem (\ref{eq:straightlinemotion})}
\begin{center}
\label{tab:Nayakshin1}
\begin{tabular}{|c|c|c|}
\hline
   Quantity & cgs & Astronomical \\
    \hline
    $u$ & 0 & 0  \\
  
    $g = -0.001 \displaystyle \frac{v^2_k}{r}$ & $-1.491 \times 10^{-6}$ & $-2.5 \times 10^{-6}$  \\
 
   $t_{stop} = \displaystyle \frac{a \rho_s}{\Sigma \Omega_K}$ & 0.986 $\times 10^3$ & $2.076 \times 10^{-4}$ \\
   
   $v_0 = 0.01 v_K$ & 6.679 $\times 10^3$ & $2.24 \times 10^{-4}$ \\
   
   $T = 2000 \pi \Omega^{-1}_K$ & 2.81 $\times 10^{12}$ & $5.619 \times 10^5$ \\ 
    
  \hline
\end{tabular}
\end{center}
\end{minipage}
\end{table*}

\subsection{Approximation Error}
\label{sec:approximation}

To understand why the numerical results for the schemes (\ref{eq:indirectZhu})-(\ref{eq:indirectSnyt}) do not coincide with the analytical results, we transformed all the test schemes to
a <<reduced>> form in which $v^{n+1}$ is a function of the values from the previous time step. The results of this transformation are presented in Table \ref{tab:scheme_approxim2}. In this form, regularization with direct operator order (\ref{eq:directZhu}) and the mixed time step scheme (\ref{eq:Nayakshin}) are identical, explaining the coincidence of the results for these computations. Moreover, the results for the schemes (\ref{eq:directZhu}) and (\ref{eq:Nayakshin}) approach the terminal velocity from the previous time step $u^n+g t_{\rm stop}$ as $t_{\rm stop} \rightarrow 0$. On the contrary, regularization with reverse operator order (\ref{eq:indirectZhu}) yields the velocity $g \tau+u^n$, and the quasianalytical approach with reverse operator order 
(\ref{eq:indirectSnyt}) the velocity $g \tau$, as $t_{stop} \rightarrow 0$. Since neither of these values depends on $t_{stop}$, the computation of the rate ends up being the same for grains with different sizes. The quasianalytical scheme with reverse operator order (\ref{eq:directSnyt}) yields velocities that depend on the grain size for small times $t_{\rm stop}$, but these values are appreciably underestimated relative to the analytical results. 

We also estimated how the error in the velocity computation in the problem (\ref{eq:straightlinemotion}) depends on the time step $\tau$ for the methods displaying high accurary in their reproduction of the body velocities. If $v^n$ is the exact solution of (\ref{eq:straightlinemotion}) we can expand $v^{n+1}$ in a Taylor series in $\tau$: 
\begin{equation}
v^{n+1}=v^n+\tau \left( \displaystyle\frac{dv}{dt} \right)^n+ 
 \displaystyle \frac{\tau^2}{2} \left( \displaystyle\frac{d^2v}{dt^2} \right)^n...+
 \displaystyle \frac{\tau^k}{k!} \left( \displaystyle\frac{d^kv}{dt^k} \right)^n...
\end{equation}
Because $v^n$ satisfies (\ref{eq:straightlinemotion}),
\begin{equation}
\left ( \displaystyle\frac{dv}{dt} \right)^n=g+\displaystyle\frac{u^n-v^n}{t_{\rm stop}}, \ \ 
\left( \displaystyle\frac{d^kv}{dt^k} \right)^n = \frac{(-1)^{k-1}}{t_{\rm stop}^{k-1}} \left ( \displaystyle\frac{dv}{dt} \right)^n,
\end{equation}
This means that the Taylor-series expansion of the solution has the form
\begin{equation}
\label{eq:TeylorEquation}
v^{n+1}=v^n+\tau\left(g+\displaystyle\frac{u^n-v^n}{t_{\rm stop}} \right)
            -\displaystyle\frac{\tau^2}{2t_{\rm stop}}
            \left(g+\displaystyle\frac{u^n-v^n}{t_{\rm stop}} \right)+O(\tau^3).
\end{equation}

Subtracting the presented numerical velocity obtained for the schemes (\ref{eq:directZhu}) and (\ref{eq:Nayakshin}) from  (\ref{eq:TeylorEquation}),
\begin{equation}
\label{eq:DerivedScheme}
v^{n+1}=(v^n\displaystyle\frac{t_{\rm stop}}{\tau}+g t_{\rm stop} + u^n)\displaystyle\frac{\tau}{\tau+t_{\rm stop}},
\end{equation}
yields the error $\varepsilon$ for the velocity
\begin{equation}
\label{eq:ErrorScheme}
\varepsilon=\displaystyle \frac{1}{2}\left(v^n-u^n-gt_{\rm stop}\right)\frac{\tau^2(\tau-t_{\rm stop})}{t^2_{\rm stop}(\tau+t_{\rm stop})}+...\frac{(-1)^{k-1} \tau^k}{k! t^k_{\rm stop}}(v^n-u^n-gt_{\rm stop})...
\end{equation}
It is clear from (\ref{eq:ErrorScheme}) that all terms in the error are
proportional to $(v^n-u^n-gt_{\rm stop})$. Since it follows from (\ref{eq:simple_analytics}) when $\displaystyle\frac{\tau}{t_{\rm stop}} \gg 1$ that $v=u+gt_{\rm stop}$, all terms in the error tend to zero. This means that the schemes (\ref{eq:directZhu}) and (\ref{eq:Nayakshin}) have infinite approximation
orders for small values of $t_{\rm stop}$. We confirmed this in computational experiments, by carrying out the integration for the problem (\ref{eq:straightlinemotion}) in the same time interval,  $[0;2000 \pi \Omega_K^{-1}]$, but with time steps of $2\tau$ and $4\tau$. We found the difference between the numerical and the analytical solutions relative to the analytical solution at the final time $T=2000 \pi \Omega_K^{-1}$. The results
for the scheme (\ref{eq:Nayakshin}) are presented in Table \ref{tab:Nayakshin}. The approximation error is close to <<machine>> accuracy,
without any increase in the error when the integration step is increased. 

The second column of Table \ref{tab:scheme_approxim2} presents the main error terms for the methods (\ref{eq:Nayakshin})-(\ref{eq:directZhu}).The main error terms for schemes (\ref{eq:indirectZhu})-(\ref{eq:indirectSnyt}) depend quadratically on $\tau$. This means that these schemes will be first-order accurate over the entire integration interval. We confirmed this using computational experiments analogous to those desribed in the previous section. The computation results for the scheme (\ref{eq:indirectZhu}) are presented in Table \ref{tab:ZhuIndir}. For bodies of all sizes in all the computations for this scheme, the error in the solution increases by a factor of two when $\tau$ is increased by this same factor. The deviation of the computational from the exact solution for these methods is inversely proportional to $t^2_{\rm stop}$; therefore, the maximum errors are expected for bodies with smaller sizes, consistent with the results presented in Fig.\ref{fig:DustyBox_splitting} and Table \ref{tab:ZhuIndir}. It follows from this table that the relative error in the solution does not exceed 10\% only for large bodies
for which $\tau \leq 0.1 t_{\rm stop}$; when $\tau =t_{\rm stop}$ the error in the velocity is 100\%. This means that, although
the schemes (\ref{eq:indirectZhu})-(\ref{eq:indirectSnyt}) do not require a restriction on the step size from the point of view of stability, and are first-order accurate in time, applying these schemes requires a constraint on the step size $\tau$ from the point of view of accuracy. The accuracy-imposed restriction $\tau \leq 0.1 t_{stop}$  for the scheme (\ref{eq:indirectZhu}) is a factor of 20 stricter than the stability-imposed restriction for the explicit method (\ref{eq:Euler}), $\tau \leq 2 t_{\rm stop}$.

\begin{table*}
\centering
\begin{minipage}{160mm}
\caption{Numerical schemes and errors in the solutions of (\ref{eq:straightlinemotion})}
  \label{tab:scheme_approxim2}
  \begin{tabular}{@{}lll@{}}
  \hline
   Method  &  Computational formula &  Error in the solution \\
     & &  \\
 \hline
   Regularization with & & \\ 
   
   direct operator & $v^{n+1}=(v^n\displaystyle\frac{t_{\rm stop}}{\tau}+g t_{\rm stop} + u^n)\displaystyle\frac{\tau}{\tau+t_{\rm stop}}$ &
   $\displaystyle \frac{1}{2}\left|v^n-u^n-gt_{\rm stop}\right|\frac{\tau^2(\tau-t_{\rm stop})}{t^2_{\rm stop}(\tau+t_{\rm stop})}$\\
   order %(\ref{eq:directZhu}) 
   & & \\
\hline

  Regularization with & & \\
  reverse operator& $v^{n+1}=(v^n\displaystyle\frac{t_{\rm stop}}{\tau}+g (t_{\rm stop}+\tau) +u^n) \displaystyle\frac{\tau}{t_{\rm stop}+\tau}$ &
   $\displaystyle \frac{1}{2}\left|(u^n-v^n)\frac{\tau-t_{\rm stop}}{\tau+t_{\rm stop}}+gt_{\rm stop}\right|\frac{\tau^2}{t^2_{\rm stop}}$\\
 order %(\ref{eq:indirectZhu})
 & & \\
 
\hline

Quasianalytical & & \\ 

integration with direct & $v^{n+1}=(v^n+(g t_{\rm stop} +u^n)\displaystyle \frac{\tau}{t_{\rm stop}})e^{-\displaystyle\frac{\tau}{t_{\rm stop}}}$& $\displaystyle\frac{1}{2}\left|g t_{\rm stop}+u^n\right|\frac{\tau^2}{t^2_{\rm stop}}^{**}$\\
 
operator order %(\ref{eq:directSnyt}) 
& & \\ 
\hline

Quasianalytical & & \\

integration with reverse & $v^{n+1}=v^n e^{-\displaystyle\frac{\tau}{t_{stop}}}+(g t_{\rm stop}+u^n)\displaystyle \frac{\tau}{t_{\rm stop}}$ & $\displaystyle\frac{1}{2}\left|g t_{\rm stop}+u^n\right|\frac{\tau^2}{t^2_{\rm stop}}^{**}$\\

operator order %(\ref{eq:indirectSnyt}) 
& & \\
\hline

Mixed layer  
      &  &
   \\
   method %(\ref{eq:Nayakshin})
   & $v^{n+1}=(v^n\displaystyle\frac{t_{\rm stop}}{\tau}+g t_{\rm stop} + u^n)\displaystyle\frac{\tau}{\tau+t_{\rm stop}}$  &  $\displaystyle \frac{1}{2}\left|v^n-u^n-gt_{\rm stop}\right|\frac{\tau^2(\tau-t_{\rm stop})}{t^2_{\rm stop}(\tau+t_{\rm stop})}$ \\
   
\hline

Quasianalytical &  & \\
integration without 
& $v^{n+1}=(g^n t_{\rm stop}+u^n)+(v^n-g^n t_{\rm stop}-u^n){\rm e}^{-\displaystyle\frac{\tau}{t_{\rm stop}}}$ & -
\\
separation %(\ref{eq:quasianalytSolution})
&  &
\\

\hline

Short friction time & & \\
approximation (\ref{eq:SFTA}) & $v^{n+1}=g^n_{rel}t_{\rm stop}+u^n$ & $>St^2 v^{n*}$ \\

\hline
\end{tabular}
$^{**}$ The error in the solution has the indicated form only when
$\displaystyle\frac{\tau}{t_{\rm stop}}<1$. \\
$^*$ Error in the solution for the radial migration of bodies (\ref{eq:system})+(\ref{eq:init}).
\end{minipage}
\end{table*}  

\begin{table*}
\centering
\begin{minipage}{140mm}
\caption{Relative error in percent in the integration for the problem (\ref{eq:straightlinemotion}) in the interval $[0;2000 \pi \Omega_K^{-1}]$ using the mixed layer scheme}
\label{tab:Nayakshin}
\begin{tabular}{@{}clllllll@{}}
\hline
   Particle size, cm & $10^{-4}$& $10^{-3}$&$10^{-2}$ & $0.1$ & $1$& $10$ & $100$  \\
    $ \ \ \ \ \ \ \ \tau / t_{stop} $ & $10^3$& $10^2$&$10$ & $1$ & $0.1$& $10^{-2}$ & $10^{-3}$  \\
    \hline
    \hline
    $\tau$ & $0$& $0$&$0$ & $1.34 \times 10^{-14}$ & $1.1\times 10^{-13}$& $8.6 \times 10^{-13}$ & $6.9 \times 10^{-12}$  \\
  
   $2\tau$ & $0$& $0$&$0$ & $0$ & $4.3\times 10^{-14}$& $4.3\times 10^{-13}$ & $3.45\times 10^{-12}$  \\
 
   $4\tau$ & $0$& $0$&$0$ & $0$ & $ 2.2\times 10^{-14}$& $2.2\times 10^{-13}$ & $1.72\times 10^{-12}$  \\
    
  \hline
\end{tabular}
\end{minipage}
\end{table*}

\begin{table*}
\centering
\begin{minipage}{140mm}
\caption{Relative error in percent in the integration for the problem (\ref{eq:straightlinemotion}) in the interval $[0;2000 \pi \Omega_K^{-1}]$ using regularization
with reverse operator order}
  \label{tab:ZhuIndir}
  \begin{tabular}{@{}clllllll@{}}
  \hline
   Particle size, cm & $10^{-4}$& $10^{-3}$&$10^{-2}$ & $0.1$ & $1$& $10$ & $100$  \\
 
  $ \ \ \ \ \ \ \ \tau / t_{stop} $ & $10^3$& $10^2$&$10$ & $1$ & $0.1$& $10^{-2}$ & $10^{-3}$  \\
   \hline
   \hline
   $\tau$   & $10^5$& $10^4$&$10^3$ & $10^2$ & $10$& $1$ & $0.1$  \\

   $2\tau $ & $2 \times 10^5$& $2 \times 10^4$&$ 2 \times 10^3$ & $ 2 \times 10^2$ & $20$& $2$ & $0.2$  \\

   $4\tau $ & $4 \times 10^5 $& $ 4 \times 10^4$&$ 4\times 10^3$ & $4 \times 10^2$ & $40$& $4$ & $0.4$  \\
    
  \hline
\end{tabular}
\end{minipage}
\end{table*}

\newpage
\section{Test 2. Simple model for radial migration of grain in a disc. Solution of a linearized system as an approximation to the solution of the original system}
\label{sec:quasianalytics}

In this section, we consider a test problem with an approximate analytical solution in which the non-aerodynamical acceleration $g$ depends non-linearly on the velocity of the body. This represents a simple model for the radial migration of bodies in a circumstellar disk. Weidenschilling 1977 \cite{Weidenschilling1977} published a qualitative analysis of the solution of this problem in 1977, which established that meter-size bodies have the maximum migration velocities toward the protostar, leading to their falling into the protostar over several hundred years. 

The equations of motion of a solid body in a disk (under the gravitational field of the star and friction with gas in the disk) written in polar coordinates have the form (e.g., \cite{Landavshitz6,ArmatageLectures2007,TakeuchiLin2002}):
\begin{equation}
\label{eq:system}
\left\{
 \begin{array}{lcl}
    
        \displaystyle 
        \frac{dr}{dt} = v_r, \\
        \displaystyle 
        \frac{dv_r}{dt} =  \frac{v_{\varphi}^2}{r} - \frac{GM}{r^2} - \frac{v_r-u_r}{t_{\rm stop}},\\
        \displaystyle
        \frac{dv_{\varphi}}{dt} = - \frac{v_r v_{\varphi}}{r} - \frac{v_{\varphi} - u_{\varphi}}{t_{\rm stop}}.
    \end{array}
\right.
\end{equation}

Here, $r,v_r,v_{\varphi}$ are the orbital radius and the radial and angular velocities of the body, $u_r,u_{\varphi}$ the radial and angular velocities of the gas, and $M$ the mass of the central body.

The following Cauchy problem can be posed for the system:

\begin{equation}
\label{eq:init}
\left\{
    \begin{array}{l}
        \left.r\right|_{t=0} = r_0, \\
        \left.v_r\right|_{t=0} = 0, \\
        \left.v_{\varphi}\right|_{t=0} = v_{K} = \sqrt{\displaystyle\frac{GM}{r}}. \\
    \end{array}
\right.
\end{equation}
Following \cite{ArmatageLectures2007}, we assumed that (1) the gas disk in which the dust particles move is in equilibrium, i.e., $u_r=0$; (2) the gas pressure decreases from the center to the periphery, i.e., the configuration of the gas disk brings about the equilibrium angular velocity $u^2_{\varphi}=(1-\eta) v^2_{\rm K}$, where $0 < \eta \ll 1$.

We took a dust particle to move in a close to Keplerian orbit and simplified the initial system (\ref{eq:system}) using the following approximations:
\begin{equation}
\label{eq:limitsKeplerian1}
\displaystyle \frac{d}{dt}(r v_\varphi) \approx v_r \frac{d}{dr}(rv_{\rm K})=\frac{1}{2}v_r v_{\rm K},
\end{equation}
\begin{equation}
\label{eq:limitsKeplerian2}
v_{\varphi}+v_{\rm K} \approx 2v_{\rm K},
\end{equation},
\begin{equation}
\label{eq:limitsKeplerian3}
\displaystyle \frac{1}{\sqrt{1-\eta}} \approx 1+\frac{\eta}{2}, \ \eta \sqrt{1-\eta} \approx \eta.
\end{equation}
Due to (\ref{eq:limitsKeplerian1}) the third differential equation in (\ref{eq:system}) becomes an algebraic equation, and the second becomes a linear equation in $v_r, \ v_{\varphi}$ by virtue of (\ref{eq:limitsKeplerian2}):

\begin{equation}
\label{eq:system2}
\left\{
 \begin{array}{lcl}
    
        \displaystyle 
        \frac{dr}{dt} = v_r, \\
        \displaystyle 
        \frac{dv_r}{dt} =  \frac{2v_{\rm K}}{r} (v_{\varphi}- v_{\rm K})- \frac{v_r}{t_{\rm stop}},\\
        \displaystyle
        \frac{1}{2}v_r v_{\rm K} = - r \frac{v_{\varphi} - u_{\varphi}}{t_{\rm stop}}.
    \end{array}
\right.
\end{equation}
Applying (\ref{eq:limitsKeplerian3}) transforms the system (\ref{eq:system2}) to the form:
\begin{equation}
\label{eq:system3}
\left\{
 \begin{array}{lcl}
        \displaystyle 
        \frac{dr}{dt} = v_r, \\
        \displaystyle 
        \frac{dv_r}{dt} = -\eta \frac{v^2_{\rm K}}{r} + \frac{2v_{\rm K}}{r} (v_{\varphi}- u_{\varphi})- \frac{v_r}{t_{\rm stop}},\\
        \displaystyle
        \frac{1}{2}v_r v_{\rm K} = - r \frac{v_{\varphi} - u_{\varphi}}{t_{\rm stop}}.
        \end{array}
\right.
\end{equation}
We can obtain a quasistationary solution for the system (\ref{eq:system3}) ((satisfying the condition $\displaystyle\frac{dv_r}{dt}=0$) if we exclude $(v_{\varphi}-u_{\varphi})$ from the second equation:
\begin{equation}
\frac{v_r}{v_{K}} = - {\eta \over St(r) + St(r)^{-1}}\;,
\label{eq:analyt}
\end{equation}

As far as we are aware, the relation (\ref{eq:analyt}) was first
presented in \cite{Nakagawa1986}. A relation for the case of a non-
equilibrium gas disk, when $u_r \neq 0$, can be found in \cite{TakeuchiLin2002}. By integration of (\ref{eq:analyt}) over the time, we obtain

\begin{equation}
\label{eq:rad_analyt_gen}
\eta t = t_{\rm stop} \ln (\displaystyle \frac{r_0}{r})+\frac{r_0^3-r^3}{3GMt_{\rm stop}},
\end{equation}

\begin{equation}
\label{eq:vphi_analyt}
v_{\varphi}=u_{\varphi}+\frac{v_r St(r)}{2}.
\end{equation}

If $t_{\rm stop} \Omega \ll 1$, Eq. (\ref{eq:rad_analyt_gen}) acquires the form
\begin{equation}
\label{eq:rad_analyt_small}
r=(r_0^3-3 GM \eta t_{\rm stop} t)^{1/3}.
\end{equation}

We do not know of any proof that the solutions of the system (\ref{eq:system}) and the transformed system (\ref{eq:system3}) are close when $\eta \ll 1$, however, our numerous numerical computations show that formulas (\ref{eq:analyt})-(\ref{eq:rad_analyt_small}) are a good approximation to the solution of the initial system (\ref{eq:system}). The similarity of the numerical solution (\ref{eq:system}) and the analytical solution (\ref{eq:analyt}) has also been shown in \cite{ChaNayakshinDust2011,NayakshinDust2017}. 

\section{Numerical solution of the non-linear system}
\label{sec:nonlinear_results}

We solved the Cauchy problem (\ref{eq:system})-(\ref{eq:init}) using the following two-step first order accurate scheme. The first step is finding the dust velocity using one of the methods (\ref{eq:Nayakshin}),(\ref{eq:quasianalytSolution}) or (\ref{eq:directZhu}). On the second step we used the resulting velocities to determine the coordinates of the particle:
\begin{equation}
\label{PosUpdate}
\left\{
 \begin{array}{lcl}
        \displaystyle 
        \frac{r^{n+1}-r^n}{\tau} =  v_r^{n+1},\\
        \displaystyle
        \frac{\varphi^{n+1}-\varphi^n}{\tau} =  v_{\varphi}^{n+1},\\
    \end{array}
\right.
\end{equation}
where $r^n,\varphi^n$ are the radius and angle of the
particle at the current time step (at time $t$) and $r^{n+1},\varphi^{n+1}$ are the coordinates of the particle at the next
time step (at time $t+\tau$).

The system (\ref{eq:system}) without terms corresponding to  the action of gas drag has the invariant $\displaystyle\frac{d(v_r^2+v_{\varphi}^2)}{dt}=2\Omega_{\rm K}(r)$. This system with the initial conditions (\ref{eq:init}) has a stationary analytical solution ($r=r_0, \ v_r=0, \ v_{\varphi}=v_{\rm K}$). It is known that the following problem arises when this system is integrated numerically (for example, using an explicit method): a diverging, spiral trajectory is obtained instead of a closed, circular orbit. Preserving the correct form of the trajectories requires reducing the time step size; the larger the number of orbits undergone by the body, the smaller the step that must be used. Simplectic schemes have been developed to avoid this restriction on the time step size while preserving the correct form of the trajectories, such as leap-frog schemes. However,
Bai and Stone \cite{BaiStone2010ApJS} showed that a simplectic leap-frog scheme requires additional constraints on the time step size in the integration of the dynamics of small bodies that are strongly tied to the gas. Thus, the choice of scheme enabling computations with the maximum time step is determined by the need to find the optimum compromise between two factors -- the requirement for the time step size from the point of view of preserving the orbit geometry and from the point of view of modeling the dynamics of small bodies. In our current study, we have solved a simpler problem. We are convinced that the use of the
scheme (\ref{eq:Nayakshin})+(\ref{PosUpdate}) with a step $\tau$ determined by the condition (\ref{eq:tauKurant}) enables preservation of the circular orbit of a body at a distance of 20~AU from the protostar over 1000 orbits, i.e., over about one million years. Further, we have compared all methods only from the point of view of the possibility of integrating the dynamics of bodies of arbitrarily small size without reducing the time step size relative to that indicated by the condition (\ref{eq:tauKurant}). 

\subsection{Reconstruction of the Velocity Field}
\label{sec:velocity_analyt}

The aim of this section is to show that a group of methods having high accuracy when solving the linear equation (\ref{eq:straightlinemotion}) also display high accuracy for the non-linear problem modeling the radial migration of bodies in a disk.

We took 20 particles for which we specified the stopping time $t_{\rm stop}=10^a \Omega^{-1}_{\rm K}$, where $a$ has values
in the interval $[-6;2]$. We placed the particles at the radius $r_0=20$~AU from a protostar with mass $M=1M_{\odot}$ in a disk with $u_{\varphi}=0.995v_{\rm K}$ and $u_r=0.$ Under these conditions, the range of Stokes numbers corresponds to bodies with radii from $10^{-6}$~m to 100~m. We numerically solved the Cauchy problem (\ref{eq:system})-(\ref{eq:init}) for each particle.
We carried out the integration over 15 orbits of the bodies at a distance of 20~AU, that is, in the interval $[0,T]=[0,30 \pi \Omega^{-1}_{\rm K}]$ with a constant time step size \textit{identical for all the bodies}, determined by relation (\ref{eq:tauKurant}). We computed the values $\displaystyle\frac{v_r}{v_{\rm K}(r)}$ for the resulting numerical solutions at $t = T$, which we then compared with the analytical values given by (\ref{eq:analyt}).

\begin{figure*}    \includegraphics[scale=0.7]{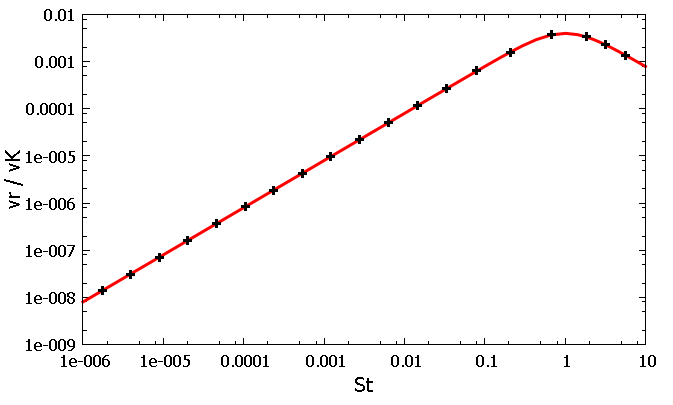} %[width=150mm]{fig2.eps}
  \caption{Solution of (\ref{eq:system})+(\ref{eq:init}) for $t=1340$~yrs using the mixed layer method (\ref{eq:Nayakshin})+(\ref{PosUpdate}). The curve shows the analytical solution (\ref{eq:analyt}) and the symbols show the numerical solution.}
\label{fig:NonlinearAnalyt}
\end{figure*}

Figure \ref{fig:NonlinearAnalyt} presents the radial drift velocity for the bodies as a function of the Stokes number computed using the mixed layer scheme (\ref{eq:Nayakshin})+(\ref{PosUpdate}). The numerical and analytical solutions coincide for all values of $\displaystyle \frac {t_{\rm stop}}{\tau} $; i.e., this method accurately conveys the loss of $\tau$ angular momentum of a body due to the gas drag, independent of the size of the body. We also obtained a similar coincidence of the numerical and analytical
results for the schemes (\ref{eq:directZhu})+(\ref{PosUpdate}) and (\ref{eq:quasianalytSolution})+(\ref{PosUpdate}). 

The dependence of the radial velocity of the bodies on the Stokes number presented in Fig. \ref{fig:NonlinearAnalyt} illustrates
the result of \cite{Weidenschilling1977} known since 1977, that the highest migration rates are displayed by bodies for which $St=1$ (under the conditions considered, these are bodies with radii of 1~m). We can also see that the radial drift speeds of meter-size bodies exceed those of micron-size particles by nearly six orders
of magnitude. This means that, in the absence of inhomogeneities in the disk that delay the rapid radial migration of meter-size bodies, such bodies will fall into the protostar on a time scale of several hundred years.

\subsection{Migration of a Ring}
\label{sec:ring_analytics}
The aim of this section is to show that a group of methods displaying high accuracy in the solution of the linear problem (\ref{eq:straightlinemotion}) correctly conveys the motion of a ring of dust particles on the scales of the disk and the characteristic time scale for the disk dynamics for the entire range of particles considered. We will also show that the schemes (\ref{eq:indirectZhu})+(\ref{PosUpdate}) and (\ref{eq:directSnyt})+(\ref{PosUpdate}) appreciably distort the migration speed of a ring when $\tau>t_{\rm stop}$.  As in Section \ref{sec:velocity_analyt} we considered a disk around a protostar with mass $M=1M_{\odot}$, and with $u_{\varphi}=0.995v_{\rm K}, \ u_r=0.$ 
We took 400 particles and distributed them uniformly in a ring [18 \textrm{AU}; 20 \textrm{AU}]. We specified the same stopping time for all the particles, which we kept constant over the entire course of the integration time:
\begin{equation}
\label{eq:InitTstopRing}
t_{\rm stop}=St \Omega^{-1}_{\rm K}(r_0),
\end{equation}
where $r_0=20$~AU, $St=2\times 10^{-3}$. According
to (\ref{eq:t_stop_sigma}) under the conditions of a massive disk described in Section \ref{sec:drag_regimes}, this stopping time corresponds to an initial radius for the bodies of $2-3$~mm. The migration of bodies in the inner region of the disk in a regime in which the stopping time remains constant implies that the radius of solid-phase particles grows in accordance with a law that depends on the surface density and radius. We carried out the integration
with a constant time step size determined by (\ref{eq:tauKurant}) over 1300 orbits of the outer boundary of the ring; i.e., $T=2600 \pi \Omega^{-1}_{\rm K}(r_0)$ or 116000~yrs.  

\begin{figure*}
  \includegraphics[scale=0.56]{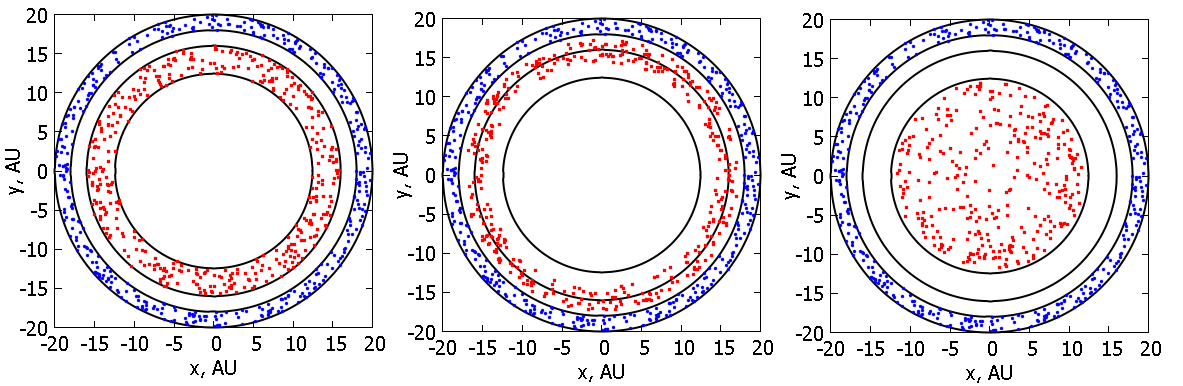} %[width=150mm]{fig2.eps}
  \caption{Motion of particles forming a ring. At the initial time, $St=2 \times 10^{-3}$, and the analytical solution is given by (\ref{eq:rad_analyt_gen}). The blue (dark gray) points show the positions of the particles at the initial time, and the red (lighter gray) points the positions of the particles after 116 000 yrs (1300 orbits). Results are shown for the mixed layer scheme (\ref{eq:Nayakshin})+(\ref{PosUpdate}) (left), quasianalytical integration with direct operator order (\ref{eq:directSnyt})+(\ref{PosUpdate}) (middle), and regularization with reverse operator order (\ref{eq:indirectZhu})+(\ref{PosUpdate}) (right); $u_{\varphi}=0.995v_K$}
\label{fig:differencesAll}
\end{figure*}

Figure \ref{fig:differencesAll} presents the results of the computations obtained using the methods (\ref{eq:Nayakshin})+(\ref{PosUpdate}),(\ref{eq:indirectZhu})+(\ref{PosUpdate}) and (\ref{eq:directSnyt})+(\ref{PosUpdate}). It is clear that the particles whose trajectories were computed using the mixed layer scheme are located inside a ring whose boundaries are in agreement with the analytical values found using (\ref{eq:rad_analyt_gen}). We obtained a similar agreement between the numerical and analytical results for the schemes (\ref{eq:directZhu})+(\ref{PosUpdate}) and (\ref{eq:quasianalytSolution})+(\ref{PosUpdate}). 

We showed in Section \ref{sec:dustybox} that quasianalytical integration with direct operator order (\ref{eq:directSnyt}) underestimates the drift velocities of small bodies. We can see in
Fig. \ref{fig:differencesAll} (middle panel) that the computed positions of the particles deviate from the exact values by more
than $2$~AU over $116000$~yrs. It is clear from Fig.\ref{fig:DustyBox_splitting} (lower left panel) that this deviation will be substantially larger for smaller bodies. Another example of a numerical artefact that arises when the scheme (\ref{eq:directSnyt}) is used is presented in \cite{BateDust2015} (Fig. 4) where the vertical settling of dust onto the plane of the disk is calculated.

On the contrary, regularization with reverse oper-
ator order (\ref{eq:indirectZhu}) substantially overestimates the migration rates of bodies. The right-hand panel of Fig.\ref{fig:differencesAll} shows that particles whose orbital radii should have been shifted from $18$ to $11$~AUpractically reached th 
protostar.

\begin{figure*}
  \includegraphics[scale=0.5]{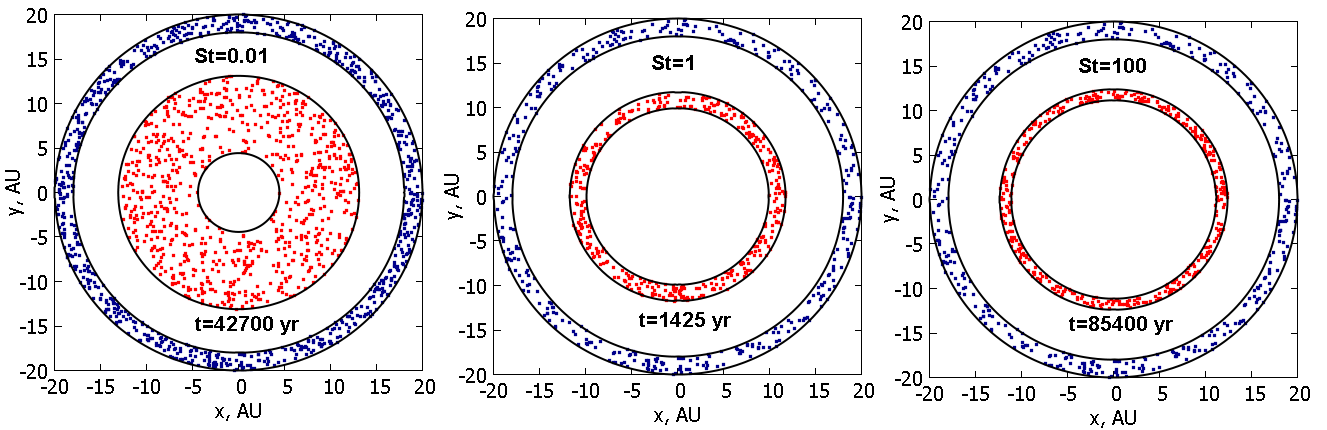} %[width=150mm]{fig2.eps}
  \caption{Motion of particles forming a ring. At the initial time, $St=10^{-2}, \ 1, \ 100$, and the size of the bodies is 0.1~mm, 1~cm, and 1~m, respectively. The curve shows the analytical solution given by (\ref{eq:rad_analyt_gen}) and the points the positions of individual particles found using the mixed layer scheme (\ref{eq:Nayakshin})+(\ref{PosUpdate}). The blue (dark gray) points show the positions of the particles at the initial time, and the red (lighter gray) points the positions of the particles after 42 700 yrs (left), 1425 yrs (middle), and 85 400 yrs (right).}
\label{fig:MixDifftstop}
\end{figure*}

Figure \ref{fig:MixDifftstop} shows the migration of a ring of particles with radii of approximately $1$~cm (left panel), $1$~m (middle panel), $100$~m (right panel) over times of $42700$, $1420$ and $85400$~yrs, respectively. The parameters of the disk are the same as in Fig. \ref{fig:differencesAll}. We carried out all these computations using the mixed layer scheme. The numerical solutions are close to the analytical solutions obtained using (\ref{eq:rad_analyt_gen}) for the entire range of particle sizes. n agreement with Fig.\ref{fig:NonlinearAnalyt} the migration speeds are maximum for meter-size particles ($St \approx 1$ for the conditions of the massive disk from Section \ref{sec:drag_regimes}). The ring expands in the course of migration toward the protostar for bodies with $St \ll 1$, and narrows for bodies with $St \gg 1$.

\section{Short Friction Time Approximation. Necessary and Sufficient Conditions for its Applicability}
\label{sec:approxBoundary}

We estimated the conditions under which the short friction time approximation correctly conveys the velocities of dust migrating in the disk along nearly Keplerian orbits.  
We applied (\ref{eq:SFTA}) to solve (\ref{eq:system})+(\ref{eq:init}) for the conditions of a gaseous disk described in Section \ref{sec:quasianalytics}, and obtained $g_{rel,r}=-\displaystyle\frac{1}{\rho}\frac{dp}{dr}$, $g_{rel,\varphi}=\displaystyle\frac{v_r v_{\varphi}}{r}$. 
Because the gaseous disk is in equilibrium with $u^2_{\varphi}=v^2_{\rm K}(1-\eta)$, we find that $-\displaystyle\frac{1}{\rho}\frac{dp}{dr}=-\eta \frac{GM}{r^2}$, so that 
\begin{equation}
\label{eq:SFTA_Kepler}
v_r=-\eta\displaystyle\frac{v^2_{\rm K}}{r} t_{\rm stop}=-\eta v_K St, \ \  v_{\varphi}=\displaystyle\frac{v_r v_{\varphi}}{r} t_{\rm stop} + v_{\rm K}\sqrt{1-\eta}.  
\end{equation}
    
On the other hand, according to (\ref{eq:analyt}) $v_r=\displaystyle - \frac{\eta v_{\rm K}} {St+St^{-1}}$. The difference in the velocities $v_r$ from (\ref{eq:SFTA_Kepler}) and (\ref{eq:analyt}) relative to the velocity (\ref{eq:SFTA_Kepler}) then acquires the form $\left|1-\displaystyle \frac{St}{\displaystyle\frac{1}{St+1/St}}  \right| = St^2$. Thus, the relation
\begin{equation}
\label{eq:SFTAlimGEN}
St^2=\varepsilon, 
\end{equation}
where $\varepsilon$ is the relative error in the computation of the radial velocity, determines a necessary condition for applicability of the short friction time approximation. That is, for particles moving in a circumstellar disk along nearly Keplerian orbits, the short friction time approximation makes it possible to obtain a solution with a relative error below 1\% ($\varepsilon<0.01$), only if 
\begin{equation}
\label{eq:SFTAlim}
St<0.1 \ \ \textrm{or} \ \ t_{\rm stop} < 0.1 \Omega^{-1}.
\end{equation}

\begin{figure*} 
\includegraphics[scale=0.55]{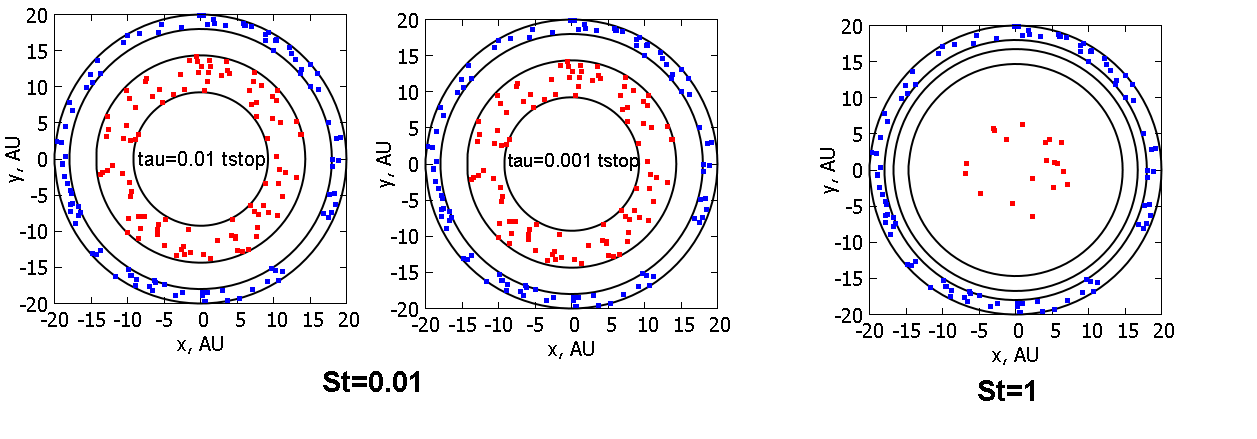} %[width=150mm]{fig2.eps}
\caption{Solution of (\ref{eq:system})+(\ref{eq:init}) using the short friction time approximation (\ref{eq:SFTA})+(\ref{PosUpdate}). The blue (dark gray) points show the positions of the particles at the initial time and the red (lighter gray) points their positions after several revolutions of the disk. In the left and middle panels, initially $St=0.01$ (0.1~mm), and the integration was carried out over $30 000$ yrs. In the right panel, initially $St=1$ (1~cm), and the integration was carried out over $450$ yrs. The time steps $\tau$ for the left and right panels were determined from the condition (\ref{eq:tauKurant}). The time step size $\tau$ for the central panel was decreased by a factor of ten relative to the value given by (\ref{eq:tauKurant}).}
\label{fig:SFTA}
\end{figure*}

We confirmed with our simulations that violation of the condition (\ref{eq:SFTAlim}) leads to appreciable deviations of the numerical from the analytical solution. Here, we computed the migration of a ring of particles similar to that described in Section \ref{sec:ring_analytics}, but using the short friction time approximation (\ref{eq:SFTA_Kepler}) instead of the mixed layer scheme (\ref{eq:Nayakshin}). We specified the stopping times of the bodies using formula (\ref{eq:InitTstopRing}), using the values $St=0.01$ and $St=1$ at $r_0=20$~AU, which corresponds to sizes of the modeled bodies from 1~cm to 1~m. We used a time step size $\tau$ determined from (\ref{eq:tauKurant}) with $r=1$~AU. The results of these computations are presented in the left and right panels of Fig.\ref{fig:SFTA}. The computed positions of particles with sizes of 1~cm ($St=0.01$, Eq. (\ref{eq:SFTAlim}) is satisfied) are located inside the ring determined by the analytical formula (\ref{eq:rad_analyt_gen}) after 30 000~yrs. The application of the short friction time approximation yields appreciably overestimated migration rates toward the center for bodies 1~m in size ($St=1$, the condition (\ref{eq:SFTAlim}) is violated). 

\begin{figure*} 
\includegraphics[scale=0.8]{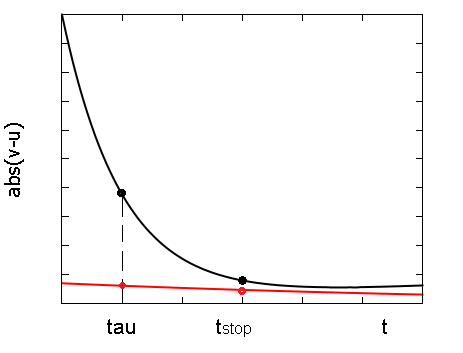} %[width=150mm]{fig2.eps}
\caption{A possible source of numerical error when $\tau<t_{stop}$ in the short friction time approximation. The upper black curve shows the case when the initial velocities of particles are appreciably different from their equilibrium values, and the lower red curve the case when the initial velocities are close to their equilibrium values.}
\label{fig:SFTA_scheme}
\end{figure*}

A substantially stricter condition for applicability of the short friction time approximation, $t_{\rm stop}<\tau$, is presented in \cite{ZhuDust} (see the text following formula (7) in that study). This condition supposes that it is possible to apply this approximation to model only particles whose sizes are a factor of 20–100 smaller than those admitted by the condition (\ref{eq:SFTAlim}). The essence of this stricter constraint is presented in Fig. \ref{fig:SFTA_scheme}, which shows the time dependences of the relative velocity between the dust and gas. If the initial velocity of the dust is far from equilibrium (black curve), and $\tau \ll t_{stop}$, the dust velocity at time $\tau$ will differ substantially from its equilibrium value. Since we simply replace the relative velocity of the dust by its equilibrium
value, $g_{rel} t_{stop}$, in the short friction time approximation, it is clear that introducing this underestimated dust velocity at earlier times can lead to appreciable deviations of the numerical from the analytical solution. In practice, this condition is not necessary if there are no mechanisms maintaining strongly non-equilibrium velocities for the dust in the system. In this case (red curve), the velocity at time $\tau$ essentially coincides with $g_{rel} t_{stop}$, so that the numerical solutions are independent of the time step size $\tau$.

We also demonstrated that this independence of the computational results on the time step size $\tau$ is indeed realized in practice, by repeating the simulations with the initial value $St=0.01$, but using $\tau=10^{-3} t_{stop}$. The results are presented in the central panel of Fig.\ref{fig:SFTA}. It is clear that the particles have moved in strict agreement with the analytical predictions.

\section{Conclusion}
\label{sec:resume}

We have considered methods for integrating the equations of dust grain motion stiffly coupled to the gas in circumstellar disk, in particular, the case when a body interacts with the gas in the Epstein (free molecular flow) regime. Various methods were compared from the point of view of their applicability for computations for bodies of arbitrary small size (from 1~$\mu$m to 10 ~m), whose velocities are mainly determined by the gas drag. Since the accurate reproduction of the trajectories of bodies with $St > 10$ over a large number of orbits requires high order accurate methods, the results presented here refer only to bodies with $St < 10$. When choosing a method for integrating the equations of motion, it is necessary to take into consideration the following factors.
\begin{itemize}
\item The explicit method \ref{eq:Euler} can be used only to compute the dynamics of large bodies, for which $\tau<2 t_{\rm stop}$. Otherwise, this scheme is unstable.

\item The fastest computational approach to modeling the radial drift of small particles is the short friction time approximation (\ref{eq:SFTA}). The relative error in the computation of the radial velocity in this method is $St^2$. This shows that this method can be applied only for bodies satisfying the condition $St<0.1$. In contrast to the earlier studies \cite{ZhuDust,BaiStone2010ApJS}, we have shown that this approach can be applied independent of the relative sizes of $\tau$ and $t_{\rm stop}$, if there are no mechanisms in the system maintaining a non-equilibrium relative velocity between the gas and bodies.

\item Several stable and fast schemes that require similar computational resources are available for the solution of the equations of motion of bodies with Stokes numbers $St < 10$. We have shown that the mixed layer scheme \ref{eq:Nayakshin}, regularization with direct operator order \ref{eq:directZhu} and quasianalytical integration \ref{eq:quasianalytSolution} accurately
reproduce the variations of the angular momenta of arbitrarily small bodies. All these methods can be used to integrate the equations of motion of bodies with time steps determined by the Courant condition for the solution of gas dynamical equations. These methods are optimal for obtaining accurate results with the
minimum number of arithmetic operations.

\item Operator splitting technique with respect to physical processes, which is often used in astrophysical codes, leads to substantial variations in the accuracy of these algorithms for small bodies, $t_{\rm stop}<\tau$, achieved in practice with the stable and fast schemes listed above. Changing the order of the operators in a scheme leads to the same result. Analysis of the errors in solutions in the linear problem DUSTYBOX and corresponding numerical tests has enabled us to determine the accuracy of schemes obtained in practice. Simulations with operator splitting tecnique with respect to physical processes satisfying the condition $t_{\rm stop}>\tau$, are admissible with any operator order and coincide with the computations without this splitting. Our analysis of the errors and test results revealed the following artefacts. Regularization with reverse operator order (\ref{eq:indirectZhu}) overestimates the angular-momentum losses for bodies whose stopping times are shorter than the time step size $\tau$. The combination of quasianalytical integration with operator splitting yields strongly distorted solutions when $\tau > t_{\rm stop}$. The angular-momentum losses are underestimated when direct operator order (\ref{eq:directSnyt}) is used, and overestimated when reverse operator order (\ref{eq:indirectSnyt}) is used.Achieving a computational accuracy of 10\% with these methods requires the use of a smaller time step size than is required to provide stability in an explicit scheme.  
\end{itemize}

\section*{Acknowledgments}
We thank S. Nayakshin, Ya. Pavlyuchenkov, V. Akimkin, and E. Lashina for discussions of this study. This work was supported by the Russian Foundation for Basic Research (grant 16-07-00916) and a grant of the President of the Russian Federation (MK-5915.2016.1), the Ministry of Education and Science of the Russian Federation (grant 3.5602.2017/BCh) and the Austrian Agency for International Mobility and Cooperation in Education, Science and Research (OEAD).

\bibliography{mybibfile}
\bibliographystyle{plain}
%\printbibliography

\end{document}